\documentclass[sigconf]{acmart}

\usepackage{amsmath,amsfonts,bm}

\def\eqref#1{equation~\ref{#1}}

\def\1{\bm{1}}

\def\ve{{\bm{e}}}

\def\vu{{\bm{u}}}
\def\vv{{\bm{v}}}

\def\mE{{\bm{E}}}

\def\mU{{\bm{U}}}
\def\mV{{\bm{V}}}

\def\mSigma{{\bm{\Sigma}}}

\DeclareMathAlphabet{\mathsfit}{\encodingdefault}{\sfdefault}{m}{sl}
\SetMathAlphabet{\mathsfit}{bold}{\encodingdefault}{\sfdefault}{bx}{n}

\usepackage{threeparttable}
\usepackage[utf8]{inputenc} 

\usepackage{booktabs}       
\usepackage{nicefrac}       
\usepackage{microtype}      
\usepackage{xcolor}         
\usepackage{soul}
\usepackage{url}
\usepackage[font={small,bf}]{caption}
\usepackage{graphicx}
\usepackage{amsmath}
\usepackage{amsthm}
\usepackage{algorithm}
\usepackage{algorithmic}
\usepackage{multirow}
\usepackage{subfigure}
\usepackage{lipsum}
\usepackage{bm}
\usepackage{color}
\usepackage{mathtools}

\usepackage{tablefootnote}

\theoremstyle{plain}

\newtheorem{theorem}{Theorem}[section]

\theoremstyle{definition}
\newtheorem{definition}[theorem]{Definition}

\theoremstyle{remark}

\AtBeginDocument{%
  \providecommand\BibTeX{{%
    \normalfont B\kern-0.5em{\scshape i\kern-0.25em b}\kern-0.8em\TeX}}}

\setcopyright{acmlicensed}
\copyrightyear{2018}
\acmYear{2018}
\acmDOI{XXXXXXX.XXXXXXX}

\acmConference[Conference acronym 'XX]{Make sure to enter the correct
  conference title from your rights confirmation emai}{June 03--05,
  2018}{Woodstock, NY}

\acmISBN{978-1-4503-XXXX-X/18/06}

\copyrightyear{2024}
\acmYear{2024}
\setcopyright{acmlicensed}\acmConference[KDD '24]{Proceedings of the 30th ACM SIGKDD Conference on Knowledge Discovery and Data Mining}{August 25--29, 2024}{Barcelona, Spain}
\acmBooktitle{Proceedings of the 30th ACM SIGKDD Conference on Knowledge Discovery and Data Mining (KDD '24), August 25--29, 2024, Barcelona, Spain} 
\acmDOI{10.1145/3637528.3671607} 
\acmISBN{979-8-4007-0490-1/24/08}

\begin{document}
\pagenumbering{arabic}
\title{Ads Recommendation in a Collapsed and Entangled World}


\author{Junwei Pan}
\affiliation{
  \institution{Tencent Inc.}
  \country{}
}
\email{jonaspan@tencent.com}

\author{Wei Xue}
\affiliation{
  \institution{Tencent Inc.}
  \country{}
}
\email{weixue@tencent.com}

\author{Ximei Wang}
\affiliation{
  \institution{Tencent Inc.}
  \country{}
}
\email{messixmwang@tencent.com}

\author{Haibin Yu}
\affiliation{
  \institution{Tencent Inc.}
  \country{}
}
\email{nathanhbyu@tencent.com}

\author{Xun Liu}
\affiliation{
  \institution{Tencent Inc.}
  \country{}
}
\email{reubenliu@tencent.com}

\author{Shijie Quan}
\affiliation{
  \institution{Tencent Inc.}
  \country{}
}
\email{justinquan@tencent.com}

\author{Xueming Qiu}
\affiliation{
  \institution{Tencent Inc.}
  \country{}
}
\email{xuemingqiu@tencent.com}

\author{Dapeng Liu}
\affiliation{
  \institution{Tencent Inc.}
  \country{}
}
\email{rocliu@tencent.com}

\author{Lei Xiao}
\affiliation{
  \institution{Tencent Inc.}
  \country{}
}
\email{shawnxiao@tencent.com}

\author{Jie Jiang}
\affiliation{
  \institution{Tencent Inc.}
  \country{}
}
\email{zeus@tencent.com}

\renewcommand{\shortauthors}{Junwei, et al.}

\begin{abstract}


We present Tencent's ads recommendation system and examine the challenges and practices of learning appropriate recommendation representations. 
Our study begins by showcasing our approaches to preserving prior knowledge when encoding features of diverse types into embedding representations. 
We specifically address sequence features, numeric features, and pre-trained embedding features.
Subsequently, we delve into two crucial challenges related to feature representation: the \emph{dimensional collapse} of embeddings and the \emph{interest entanglement} across different tasks or scenarios. 
We propose several practical approaches to address these challenges that result in robust and disentangled recommendation representations.
We then explore several training techniques to facilitate model optimization, reduce bias, and enhance exploration. 
Additionally, we introduce three analysis tools that enable us to study feature correlation, dimensional collapse, and interest entanglement.
This work builds upon the continuous efforts of Tencent's ads recommendation team over the past decade. 
It summarizes general design principles and presents a series of readily applicable solutions and analysis tools. The reported performance is based on our online advertising platform, which handles hundreds of billions of requests daily and serves millions of ads to billions of users.
  
\end{abstract}



\maketitle

\section{Introduction}
\label{sec:intro}

The online advertising industry, valued at billions of dollars, is a remarkable example of the successful application of machine learning. 
Various advertising formats, including sponsored search advertising, contextual advertising, display advertising, and micro-video advertising, heavily rely on the accurate, efficient, and reliable prediction of ads click-through or conversion rates using learned models.

Over the past decade, deep learning has achieved remarkable success in diverse domains, including computer vision (CV)~\cite{AlexNet2012, resnet2016}, natural language processing (NLP)~\cite{Transformer2017, devlin2018bert, achiam2023gpt}, and recommendation systems~\cite{DL4CTR2021, DMLR2022}. 
The effectiveness of deep learning critically depends on the selection of appropriate data representations~\cite{RepresentationLearningPerspective2013, DisentangledRepresentationLearning2022, UnderstandingCLfromAlignmentAndUniformity2020}.
Researchers have extensively explored various aspects of representation learning in CV and NLP. 
These investigations have focused on topics such as priors~\cite{Transformer2017}, smoothness and the curse of dimensionality~\cite{bengio2005curse}, depth and abstraction~\cite{bengio2011expressive}, disentangling factors of variations~\cite{DisentangledRepresentationLearning2022}, and the uniformity of representations~\cite{UnderstandingDimensionCollapse2021, hua2021feature}.

\begin{figure*}[!tp]
    \centering
    \includegraphics[width=0.95\linewidth]{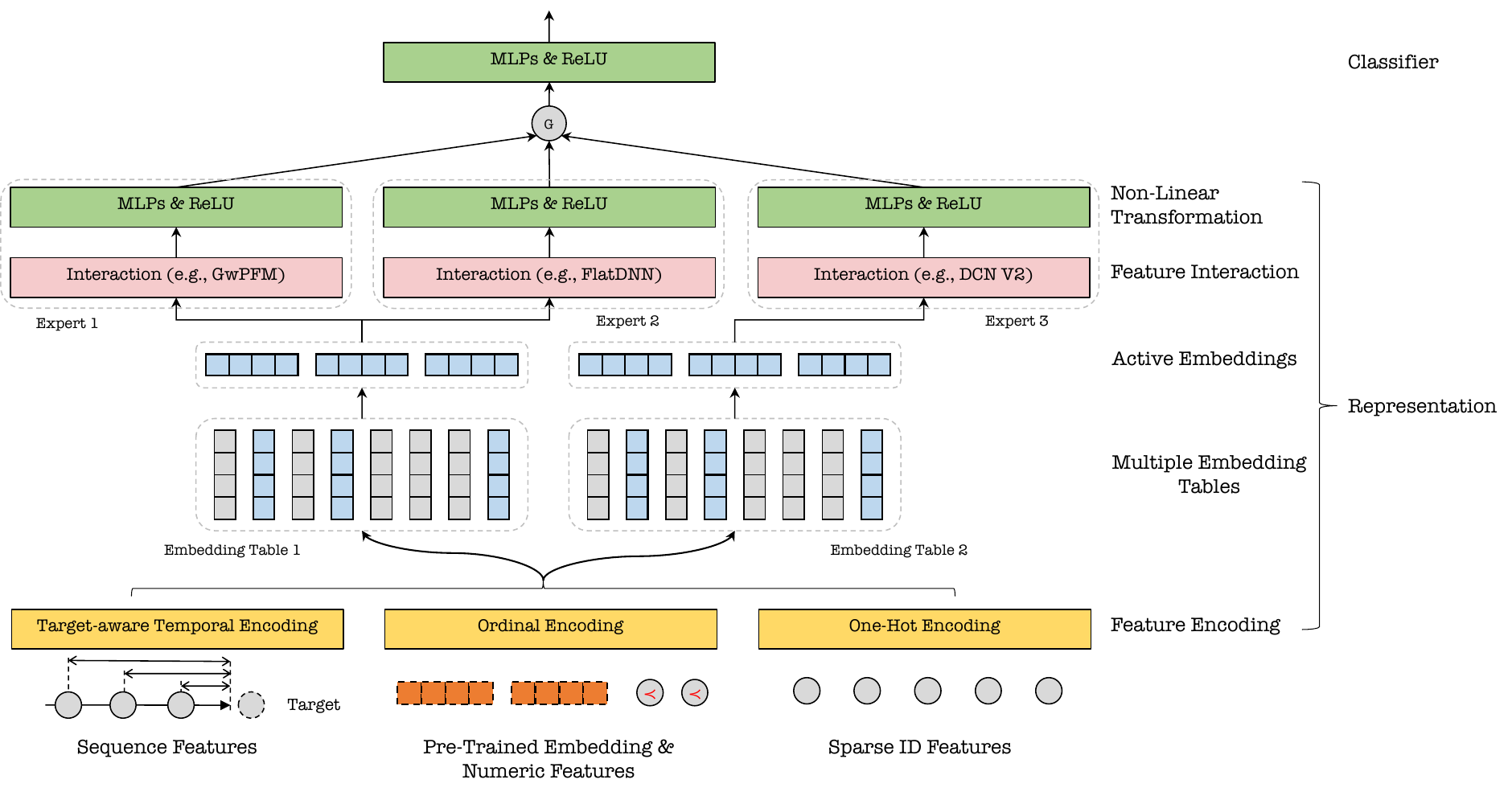}
    \caption{Architecture of our Heterogeneous Mixture-of-Experts with Multi-Embedding for single-task learning, which consists of four key modules: feature encoding, multi-embedding lookup, experts (feature interactions and MLPs), and classification towers.}
    \label{fig:overall_architecture}
\end{figure*}

In the field of recommendation systems, numerous works have focused on representation learning techniques to handle various types of features~\cite{DIN2018, SASRec2018, AutoDis2021, N-ary2022, LCRec2023}, capture feature correlations through explicit or implicit feature interactions~\cite{WideAndDeep2016, DeepFM2017, PNN2016, xDeepFM2018, DCNv22021, FinalMLP2023, tian2023eulernet}, address the entangled interest within users' complex behaviors~\cite{wang2022disentangled}, particularly in multi-task~\cite{MMoE2018, PLE2020} or multi-scenario~\cite{Star2021, PEPNet2023, HiNet2023} settings, and enhance data representation through self-supervised learning~\cite{s3-rec2020, wang2023cl4ctr}.
Despite the progress made in these representation-oriented works, several fundamental questions regarding representation learning in large-scale real-world ads recommenders remain unanswered. 

\vspace{-6pt}
\begin{itemize}
    \item \emph{Priors for Representation:} Real-world systems encompass various types of features from diverse sources, including sequence features (\textit{e.g.}, user click/conversion history), numeric features (\textit{e.g.}, semantic-preserving ad IDs), and embedding features from pre-trained external models (\textit{e.g.}, GNN or LLM). 
    Preserving the inherent priors of these features when encoding them in recommendation systems is crucial.
    \item \emph{Dimensional Collapse:} The encoding process maps all features into embeddings, typically represented as $K$-dimensional vectors, and are learned during model training. 
    However, we observe that the embeddings of many fields tend to occupy a lower-dimensional subspace instead of fully utilizing the available $K$-dimensional space. 
    Such dimensional collapse not only leads to parameter wastage but also limits the scalability of recommendation models.
    \item \emph{Interest Entanglement:} User responses in ads recommender systems are determined by complex underlying factors, particularly when multiple tasks or scenarios are learned simultaneously. 
    Existing shared-embedding approaches~\cite{SharedBottom1997, MMoE2018, PLE2020} may fail to disentangle these factors adequately, as they rely on a single entangled embedding for each feature.
\end{itemize}

This paper presents our practices for addressing these challenges. The remaining sections of the paper are organized as follows:
Section~\ref{sec:overview} provides an overview of our model architecture, giving a high-level understanding of the system.
Section~\ref{sec:feature_encoding} focuses on the encoding techniques used to integrate temporal, ordinal, and distance priors of different feature types into the representation. 
Section~\ref{sec:feature_interaction} delves into the root causes of the embedding dimensional collapse and proposes several solutions to mitigate this issue. 
Section~\ref{sec:MTL} explores the challenge of interest entanglement across various tasks and scenarios and our solutions.
Section~\ref{sec:training} presents various model training techniques. 
Finally, Section~\ref{sec:tools} introduces a set of off-the-shelf tools designed to facilitate the analysis of feature correlations, dimensional collapse, and interest entanglement.
Due to space limitations, this paper cannot provide a detailed description of each approach. For more in-depth information, please refer to the corresponding paper cited in each section.

\section{Brief System Overview}
\label{sec:overview}

The overall architecture of our ads recommendation model for single-task learning is illustrated in Fig.~\ref{fig:overall_architecture}. 
For the multi-task learning model architecture, please refer to Fig.~\ref{fig:stem_archi}.
Our model follows the widely adopted Embedding \& Explicit Interaction framework~\cite{DL4CTR2021, DMLR2022}, which consists of four key modules: feature encoding, multi-embedding lookup, experts (feature interactions and MLPs), and classification towers.
In the feature encoding module, we apply specific encoding methods tailored to various feature types in our system.
Next, based on the encoded IDs obtained from the feature encoding module, multiple embeddings are looked up from individual embedding tables for each feature. 
Within the expert module, embeddings from the same table are explicitly interacted with one another.
The outputs of the expert module are then passed through Multi-Layer Perceptrons (MLPs) with non-linear transformations.
The classification towers receive the gate-weighted sum of the outputs from the experts. Finally, the sigmoid activation function is applied to generate the final prediction.

In the case of single-task learning, such as Click-Through Rate (CTR) prediction, our model employs a single tower, as depicted in Fig.~\ref{fig:overall_architecture}. 
However, in the context of multi-task learning (MTL), such as Conversion Rate (CVR) prediction, where each conversion type is treated as an individual task~\cite{MT-FwFM2019}, our model utilizes multiple towers and corresponding gates. 
Each tower is dedicated to a specific group of conversion types, allowing for task-specific predictions.
To address the challenge of interest entanglement that arises in the MTL setting, further evolution of the model architecture to disentangle user interest is presented in Section~\ref{sec:MTL}.

Our team is responsible for ads recommendation across all modules, including retrieval and pre-ranking, CTR prediction (pCTR), (shallow) conversion prediction (pCVR) of various conversion types, deep conversion prediction (pDCVR), and Long-time Value prediction (pLTV). 
There are many commonalities regarding the model design principle among these modules, and we mainly discuss the pCTR and pCVR as representative modules for single-task and multi-task learning, respectively.
Our models serve various ads recommendation scenarios within Tencent, encompassing Moments (social stream), Channels (micro-video stream), Official Accounts (subscription), Tencent News, Tencent Video (long-video platform), and Demand Side Platform.

\begin{figure*}[h]
    \centering
    \captionsetup{labelfont=bf}
    \subfigure[\centering Temporal Interest Module]{
        \begin{minipage}[b]{0.42\linewidth}
    	\label{subfig:TIM}
  		\includegraphics[width=1\linewidth]{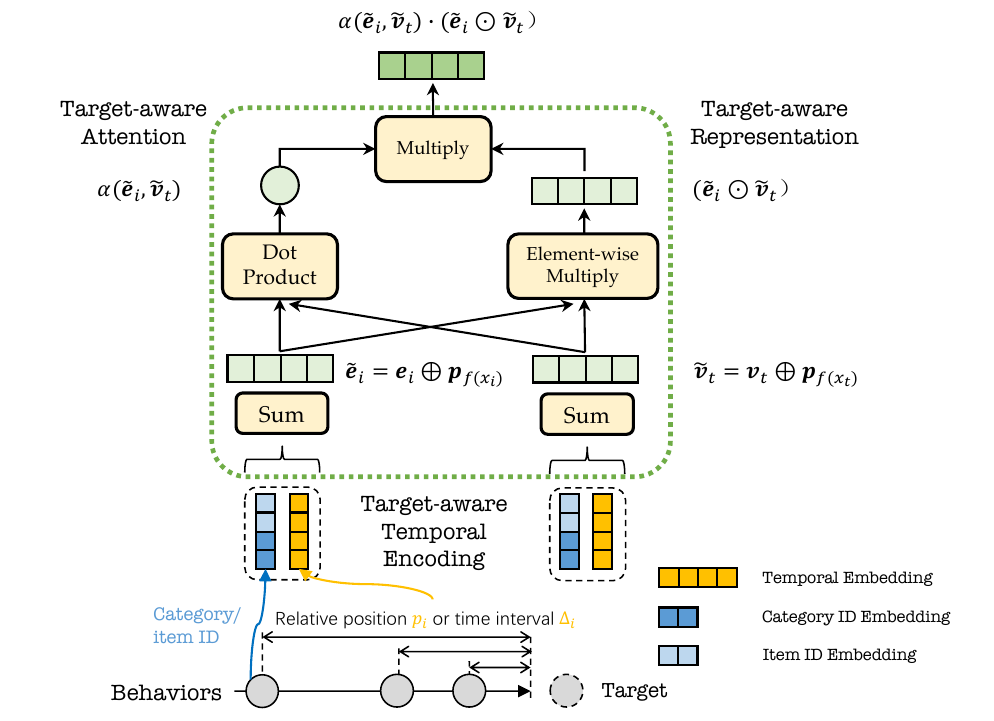}
        \end{minipage}}
    \subfigure[\centering Multiple Numeric Systems Encoding]{
		\begin{minipage}[b]{0.56\linewidth}
			\label{subfig:EE}
            \includegraphics[width=1\linewidth]{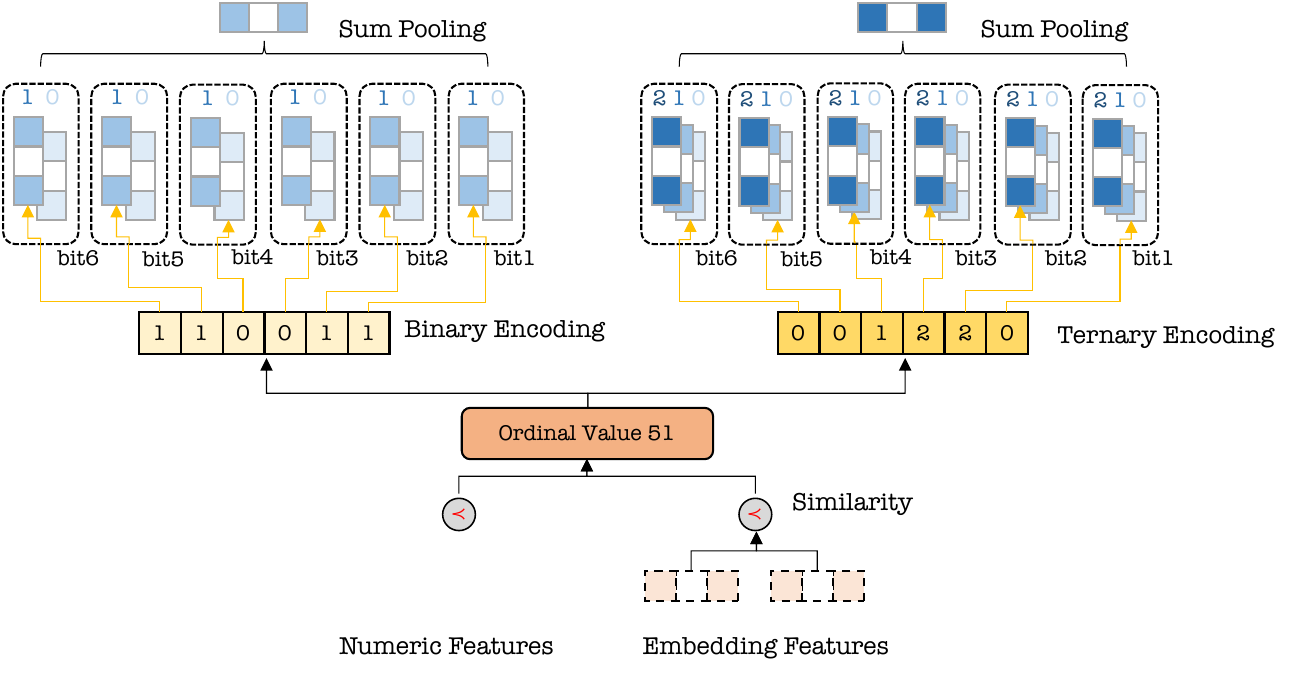}
	\end{minipage}}
    \caption{Illustration of Temporal Interest Module (left) for sequence features and Multiple Numeral Systems Encoding (right) for numeric and pre-trained embedding features.}
    \label{fig:GNN_embedding}
\end{figure*}

\vspace{-3pt}
\section{Feature Encoding}
\label{sec:feature_encoding}

In industrial ads recommendation systems, features are generated from many sources and belong to different types, such as sequence, numeric, and embedding features.
When encoding these features, we'd like to preserve their inherent temporal, ordinal, or distance (similarity) priors as much as possible.

\subsection{Sequence Features}

A user's history behaviors reflect her interest, making them critical in recommendations. 
One key characteristic of such features is that there are strong semantic and temporal correlations between these behaviors and the target~\cite{TIN2023}.
For example, given a target ad, those behaviors that are either semantically related (\textit{e.g.}, belonging to the same category with the target ad) or temporally close to the target are more informative to predict the user's response to the target item.

We propose Temporal Interest Module (TIM)~\cite{TIN2023} to learn the quadruple semantic-temporal correlation between \emph{(behavior semantic, target semantic, behavior temporal, target temporal)}.
Specifically, in addition to the semantic encoding~\cite{DIN2018, DIEN2019, DSIN2019}, TIM leverages Target-aware Temporal Encoding for each behavior, \textit{e.g.}, the \emph{relative position} or \emph{time interval} between each behavior and target.
Furthermore, to capture the \emph{quadruple correlation}, TIM employs Target-aware Attention and Target-aware Representation to interact behaviors with the target in both attention and representation, resulting in \emph{explicit 4-way interaction}(shown in Fig.~\ref{subfig:TIM}). 
Mathematically, the encoding of user behavior sequence $\mathcal{H}$ can be formulated as:

\begin{equation}
    \bm{u}_{\text{TIM}} = \sum_{X_i \in \mathcal{H}}  \alpha(\tilde{\bm{e}}_{i}, \tilde{\bm{v}}_{t}) \cdot (\tilde{\bm{e}}_{i} \odot \tilde{\bm{v}}_t)
\end{equation}

where $\alpha(\tilde{\bm{e}}_{i}, \tilde{\bm{v}}_{t})$ denotes the \emph{target-aware attention} between each behavior $i$ and target $t$,  $(\tilde{\bm{e}}_{i} \odot \tilde{\bm{v}}_t)$ denotes the \emph{target-aware representation}, and $\tilde{\bm{e}}_i = \bm{e}_i \oplus \bm{p}_{f(X_i)}$ denotes the \emph{temporally encoded embedding} of the $i$-th behavior, which is an element-wise summation of semantic embedding $\bm{e}_i$ and target-aware temporal encoding $\bm{p}_{f(X_i)}$, \textit{i.e.}, the embedding of either the relative position of each behavior regarding the target, or the discretized time interval. 
Please note that the target-aware representation $\tilde{\bm{e}}_{i} \odot \tilde{\bm{v}}_t$ acts like a feature interaction layer to explicitly interact the behavior feature with the target, as done in other FM-based explicit feature interaction models~\cite{FM2010, FFM2016, FwFM2018, xDeepFM2018, DCNv22021}.
The importance of such explicit behavior-target interaction in the representation was also emphasized in a recent work~\cite{hstu2024}.

\paragraph{Deployment Details}
In practice, we adopt both relative position and time interval for temporal encoding.
The output of TIM is concatenated with the output of the feature interaction module, \textit{e.g.}, DCN V2~\cite{DCNv22021} or GwPFM (will be discussed later).
We apply TIM on the user's click/conversion category sequence features in various click and conversion prediction tasks across multiple scenarios.
TIM brings a $1.93\%$ Gross Merchandise Value (GMV) lift in WeChat pCTR and a $2.45\%$ GMV lift in Game and e-Commerce pLTV.
We observe the model learns \emph{much stronger decaying patterns in the time interval embeddings than the relative position embedding}.
This is because users' clicks on ads are pretty sparse, making time intervals more informative than relative positions.

\subsection{Numeric Features}

Unlike independent ID features, there is inherent partial order between numeric/ordinal features, such as \text{Age\_20} $\prec$ \text{Age\_30}. 
To preserve these ordinal priors, we adopt a simplified variant of the NaryDis encoding~\cite{N-ary2022}, namely the Multiple Numeral Systems Encoding (MNSE). 
It encodes numeric features by getting codes according to multiple numeral systems (\textit{i.e.}, binary, ternary, decimal) and then assigns learnable embeddings to these codes, as shown in Fig.~\ref{subfig:EE}.

For example, a feature value "51" is transformed into code "\{6\_1, 5\_1, 4\_0, 3\_0, 2\_1, 1\_1\}" according to binary system, and "\{6\_0, 5\_0, 4\_1, 3\_2, 2\_2, 1\_0\}" according to ternary system.
All codes are projected to embeddings and then sum-pooled to get the final encoding result.
To improve computation efficiency, we remove the inter- and intra-attention in the original NaryDis~\cite{N-ary2022}.
Given a continuous feature with value $v$, the encoding result is:

\begin{equation}
    \begin{aligned}
     f_\text{MNS}(v) & = [\sum_{k=1}^{K_2} \mathbf{X}_{2k + \mathbb{B}_k}^{(2)}, \sum_{k=1}^{K_3} \mathbf{X}_{3k + \mathbb{C}_k}^{(3)} , \dots , \sum_{k=1}^{K_n} \mathbf{X}_{nk + \mathbb{N}_k}^{(n)}] \\ 
     \mathbb{B} &  = {\rm func}\_{\rm binary}(v), \ 
    \mathbb{C} = {\rm func}\_{\rm ternary}(v), \dots
\end{aligned} 
\label{eq:nEE_equation}
\end{equation}
where $\mathbf{X}_{2k + \mathbb{B}_k}^{(2)}$ and $\mathbf{X}_{3k + \mathbb{C}_k}^{(3)}$ are the embedding dictionaries for binary and ternary systems respectively, whose lengths of encodings are $K_2$ and $K_3$. 
${\rm func}\_{\rm binary}$ and ${\rm func}\_{\rm ternary}$ are the binarization and ternarization functions that transform the continuous feature $v$ into their corresponding encodings.

\paragraph{Deployment Details}

In an advertising system, ads are often indexed by discrete identifiers (Ad IDs), which are self-incremental or random and contain little information. 
However, each ad is associated with a creative containing abundant visual semantics.
We replace the self-incremental or random Ad IDs with novel Visual Semantic IDs to preserve the visual similarity between ads. 
We achieve this by obtaining visual embeddings from ad images using a vision model and applying hashing algorithms like Locality-Sensitive Hashing (LSH)~\cite{LSH} to preserve visual similarity. 
The Visual Semantic IDs serve as numeric features, and we apply Minimum Norm Scaling (MNS) to preserve their ordinal priors. This replacement leads to a 1.13\% GMV lift in Moments pCVR, with a larger lift of 1.74\% for new ads. 
Additionally, the coefficient of variation in prediction scores among similar ads exposed to the same user is significantly reduced from 2.44\% to 0.30\%, substantiating that our approach can preserve the visual similarity priors.

\vspace{-3pt}
\subsection{Embedding Features}

Besides the main recommendation model, we may train a separate model, such as LLM or GNN, to learn embeddings for entities (users or items).
Such embeddings capture the relationship between users and items from a different perspective, \textit{e.g.}, a Graph Model or a Self-Supervised Language Model, and can be trained on a larger or different dataset and hence should provide extra information to the recommendation models.
The key challenge in leveraging such pre-trained embedding directly in our recommendation system is the \textit{semantic gap} between the embedding space of the external models and the recommenders.
That is, these embedding captures different semantics from the collaborative semantics of the ID embeddings in recommendation models~\cite{LCRec2023, CTRL2023}.

We propose a Similarity Encoding Embedding approach to mitigate such a semantic gap. 
Take GNN for example. 
Once we train a GNN model and get the pre-trained embeddings $\bar{\bm{e}}_u, \bar{\bm{e}}_i$ for each user-item pair $(u, i)$, we first calculate the similarity $w_\text{sim}(u, i)$ based on their GNN embeddings using the corresponding similarity function, \textit{i.e.}, cosine in GraphSage~\cite{GraphSage2017}. 
Formally,

\vspace{-3pt}
\begin{equation} 
    w_\text{sim}(u, i) = \text{sim}(\bar{\bm{e}}_u, \bar{\bm{e}}_i).
\end{equation}

Such a similarity score is an ordinal value. 
Hence, similar to numeric features, we can use the Multiple Numeral Systems Encoding mentioned before to transform it into a learnable embedding $\bm{e}_\text{sim}(u,i) = f_\text{MNS}(w_\text{sim}(u,i))$.
After that, the encoded embedding is simultaneously co-trained with the other ID embeddings in recommenders. 
Thus, the similarity priors in the original space are retained via the similarity score and encoding. 
Then, such priors are transferred to the recommenders by aligning the similarity encoding embedding $\bm{e}_\text{sim}(u,i)$ with the recommendation ID embeddings.

Furthermore, such an embedding encoding strategy has also been developed to incorporate large-language model (LLM) knowledge into our recommendation system. 
An LLM model is first transformed into an encoder-only architecture and trained with base proxy tasks like next-sentence prediction. Through such general pre-training, the LLM encoder can encode semantic embedding. 
After that, the LLM model is finetuned with high-quality positive and negative user-item pairs from the ads domain. Such contrastive alignment enables the LLM to generate high-quality pre-trained user embeddings $\bar{\bm{e}}_u$ and ad embeddings $\bar{\bm{e}}_i$. 
With such LLM similarity priors, like GNN embeddings, we can then adopt Similarity Encoding Embedding for space alignment.

\paragraph{Deployment Details}

We train a GraphSage~\cite{GraphSage2017} upon a user-ad/content bipartite graph, with clicks in both ad and content recommendation domains as the edges. 
We then adopt the Similarity Encoding Embedding on the GNN embeddings and concatenate the resulting representation with that of the feature interaction layer.
GNN embeddings are successfully deployed in many scenarios, leading to +1.21\%, +0.59\%, and 1.47\% GMV lift on Moments, Channel, and Applet pCTR. 
In addition, incorporating LLM also leads to +2.55\% GMV lift on Channel pCVR and +1.41\% GMV lift on Channel pCTR during online A/B test.

\section{Tackling Dimensional Collapse}
\label{sec:feature_interaction}

After encoding, all features are transformed into embeddings and then interact with each other explicitly through FM-like models~\cite{MF2009, FM2010, FFM2016, DeepFM2017, PNN2016, FwFM2018, FmFM2021, xDeepFM2018, DCNv22021}.
However, one key side effect of explicit feature interaction is that some dimensions of embeddings collapse~\cite{ME2023}. 
In this section, we'll first explain the dimensional collapse phenomenon and then present two different multi-embedding approaches and a collapse-resilient feature interaction function to mitigate it.

\subsection{Embedding Dimensional Collapse}

Recent work ~\cite{achiam2023gpt, zhao2023survey, gong2023multimodal} has demonstrated that large-scale models especially transformer-based models with billions, even trillions, of parameters can achieve remarkable performance (\textit{e.g.}, GPT-4~\cite{achiam2023gpt}, LLaMA~\cite{touvron2023llama}).
Inspired by these works, we explore how to scale up ads recommendation models.
Usually, embeddings dominate the number of model parameters. 
For example, more than 99.99\% of parameters in our production model are from feature embeddings.
Therefore, we start to scale up our model by enlarging the embedding size $K$, \textit{e.g.}, increasing $K$ from $64$ to $192$.
However, it doesn't bring significant performance lift and sometimes even leads to performance deterioration.

We investigate the learned embedding matrix of each field by singular spectral analysis~\cite{UnderstandingDimensionCollapse2021}, and observe dimensional collapse.
That is, many singular values are very small, indicating that embeddings of many fields end up spanning a lower-dimensional subspace instead of the entire available embedding space~\cite{ME2023, hua2021feature}.
The dimensional collapse of embeddings results in a vast waste of model capacity since many embedding dimensions collapse and are meaningless.
Furthermore, the fact that many embeddings have already collapsed makes it infeasible to scale up models by simply increasing dimension length~\cite{ardalani2022understanding, ME2023}.

We study the root cause of the dimensional collapse and find it's due to the explicit feature interaction module, namely, fields with collapsed dimension make the embeddings of other fields collapse.
For example, some fields such as Gender have very low cardinality $N_\text{Gen}$, making their embeddings only able to span a $N_\text{Gen}$-dimension space.
As $N_\text{Gen}$ is much smaller than embedding size $K$, the interaction between these low-dimension embeddings and the possibly high-dimensional embedding (in $K$-dimensional) of remaining fields make the latter collapse to an $N_\text{Gen}$-dimensional subspace.

\subsection{Multi-Embedding Paradigm}

We propose a \emph{multi-embedding paradigm}~\cite{ME2023} to mitigate embedding dimensional collapse when scaling up ads recommenders.
Specifically, for each feature, instead of looking up only one embedding in the existing single-embedding paradigm, we learn multiple embedding tables, and look up several embeddings from these table for each feature.
Then, all feature embeddings from the same embedding table interact with each other in the corresponding expert $I$.
Formally, a recommendation model with $T$ embedding tables is defined as:


\vspace{-10pt}
\begin{align*}
    \ve_i^{(t)}&=\left(\mE_i^{(t)}\right)^{\top}\1_{x_i},\ \forall i\in\{1,2,...,N\}, \\
    h^{(t)}&=I^{(t)}\left(\ve_1^{(t)},\ve_2^{(t)},...,\ve_N^{(t)}\right), \\
    h&=\frac1T\sum_{t=1}^T g^{(t)} h^{(t)},\quad \hat{y}=F(h),
\end{align*}

where $t$ stands for the index of the embedding table, $g$ denotes the gating function for each expert, and $F(\cdot)$ denotes the final classifier.
One requirement is that there should be non-linearities such as ReLU within the interaction expert $I$; otherwise, the model is equivalent to the single-embedding paradigm~\cite{ME2023}. 
An overall architecture is shown in Figure~\ref{fig:overall_architecture}. 





The multi-embedding paradigm offers an effective approach to scaling up recommendation models. 
\emph{Instead of simply increasing the length of a shared embedding for each feature, this paradigm involves learning multiple embeddings for each feature}.
By adopting the multi-embedding paradigm, we can achieve \emph{parameter scaling} for recommendation models, which has traditionally been a challenging task~\cite{ardalani2022understanding}: the model's performance improves as the number of parameters increases.




\paragraph{Deployment Details}
Almost all pCTR models in our platform adopt the Multi-Embedding paradigm.
Specifically, we learn multiple different feature interaction experts, \textit{e.g.}, GwPFM (a variant of FFM, which will be described below), IPNN, DCN V2, or FlatDNN, and multiple embedding tables.
One or several experts share one of these embedding tables.
We name such architecture \emph{Heterogeneous Mixture-of-Experts with Multi-Embedding}, which differs from DHEN~\cite{zhang2022dhen} in the sense that~\cite{zhang2022dhen} employs one shared embedding table while we deploy multiple ones.
For example, the Moments pCTR model consists of a GwPFM, IPNN~\cite{PNN2016}, FlatDNN, and two embedding tables.
GwPFM and FlatDNN share the first table, while IPNN uses the second one.
Switching from a single embedding to the above architecture brings a $3.9\%$ GMV lift in Moments pCTR, which is one of the largest performance lifts during the past decade.

\subsection{GwPFM: Yet Another Simplified Approach to Multi-Embedding Paradigm}

FFM~\cite{FFM2016} can also be regarded as another approach of the Multi-Embedding paradigm because FFM also learns multiple embeddings for each feature.
In particular, for a dataset with $M$ fields, FFM learns $M-1$ embeddings $\{\bm{e}_{i,F_l} | F_l \neq F(i)\}$ for each feature $x_i$.
When interacting feature $x_i$ with another feature $j$, among $x_i$'s embeddings, FFM chooses the one corresponding to the field of $j$, \textit{i.e.}, $\bm{e}_{i, F(j)}$, where $F(j)$ denotes the field of feature $j$.

Even though FFM has been proven more effective than FM, it's not widely deployed in industry due to its huge space complexity since it introduces $M-2$ times more parameters than FM, where  $M$ is usually at the magnitude of hundreds in practice.
To tackle the high complexity, we propose to decouple the number of embeddings per feature from the number of fields.
Specifically, we group fields to $P$ \emph{field parts} and learn $P$ embeddings for each feature, one for each field part.
We choose a small $P$ to reduce the total model size.
Furthermore, we want to capture the field-pair-wise correlation to improve performance~\cite{FwFM2018}.
A straightforward implementation is to assign a weight for each field pair, but it leads to a computation cost of $O(M^2)$, which is unacceptable.
To reduce the computation cost,  we group fields into \emph{field groups} and learn a weight for each field group pair.
We name this method Group-weighted Part-aware Factorization Machines, or GwPFM in short.
The formal representation of its interaction is:

\begin{equation}
    \Phi = \sum^{\oplus}_{i=1} \sum^{\oplus}_{j=i+1} x_{i} x_{j} \langle \bm{e}_{i, P(j)}, \bm{e}_{j, P(i)} \rangle r_{G(i), G(j)}
\end{equation}

where $\oplus$ denotes element-wise summation, $P(i)$ and $G(i)$ denotes the field Part or Group of feature $i$, and $r_{G(i), G(j)}$ denotes the learnable weights for field group pair $(G(i), G(j))$. 

\paragraph{Deployment Details}
In practice, we split all fields into two parts: the first one consists of all fields unrelated to the target ads, including all user and context-side fields, while the second part consists of all fields regarding the target ad.
We then split all fields of the first part into $G$ groups based on expert knowledge, where $G$ is at the dozens level, usually less than 50. 
We don't further split fields in the second part to avoid high online inference complexity. 
That is, all fields in the second part belong to one field group.
The GwPFM has been deployed in our production since 2018 and has served many modules and scenarios to this day.


\subsection{Beyond Multi-Embedding Paradigm: Collapse Resilient Feature Interaction}

In addition to exploring multiple embeddings, we have conducted further investigation into the interaction function between two feature embeddings.
The conventional approach, as employed in FM, conducts an element-wise inner product between the two embeddings: 
$f(\bm{e}_i, \bm{e}_j) = \bm{e}_i \odot \bm{e}_j$.
However, recent research~\cite{UnderstandingDimensionCollapse2021} has revealed that directly calculating the distance between two embeddings can lead to dimensional collapse. 
To address this issue, researchers verify that adding a projection matrix upon embeddings before computing the inner product can effectively mitigate the collapse~\cite{UnderstandingDimensionCollapse2021}.
We confirm its efficacy in ads recommendation, that is, incorporating a field-pair wise projection matrix $M_{F(i) \to F(j)} \in \mathcal{R}^{K \times K}$ within feature interaction~\cite{FibiNet2019, DCNv22021, FmFM2021, ME2023} as done in FiBiNET, FmFM and DCN V2 can also mitigate the dimensional collapse of embeddings in recommendation. 
Formally, the interaction function between feature embedding pair $(\bm{e}_i, \bm{e}_j)$ with a projection matrix $M_{F(i) \to F(j)}$ is defined as 

\begin{equation}
    f_\text{Proj}(\bm{e}_i, \bm{e}_j) = (\bm{e}_i M_{F(i) \to F(j)}) \odot \bm{e}_j
\end{equation}

\section{Tackling Interest Entanglement}
\label{sec:MTL}


\begin{figure}[t!]
	\centering  
	\subfigbottomskip=0pt 
	\subfigcapskip=0pt 
	\subfigure[Embeddings of Single Task Like]{
		\includegraphics[width=0.45\linewidth]{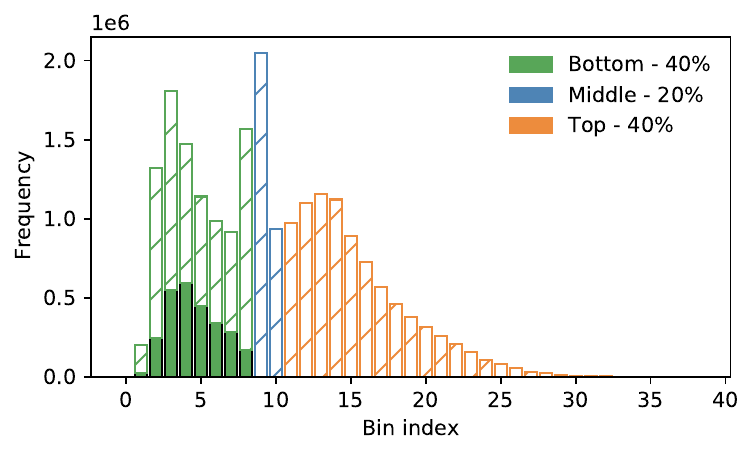}
            \label{subfig:st_like_distance_distribution}}
	\subfigure[Embeddings of Single Task Finish]{
		\includegraphics[width=0.45\linewidth]{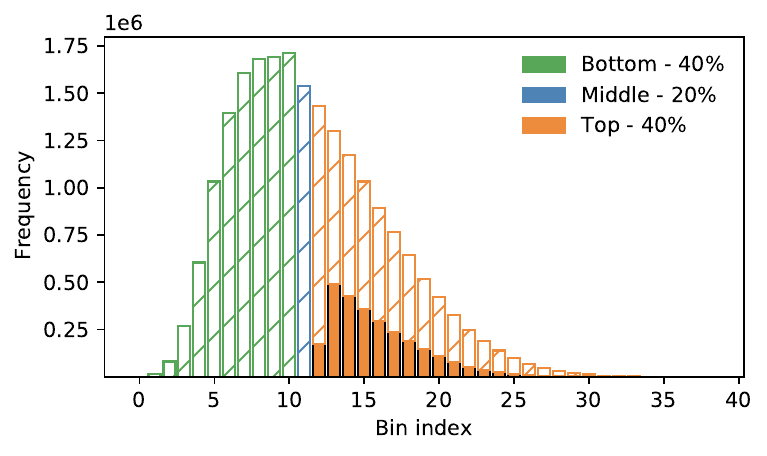}
            \label{subfig:st_finish_distance_distribution}}
	\subfigure[Shared-Embeddings in PLE]{
		\includegraphics[width=0.45\linewidth]{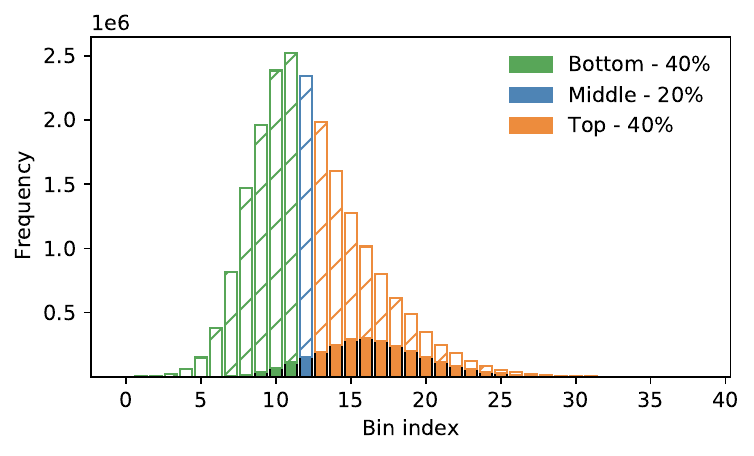}
            \label{subfig:ple_distance_distribution}}
	\subfigure[Like-Specific Embeddings in STEM]{
		\includegraphics[width=0.45\linewidth]{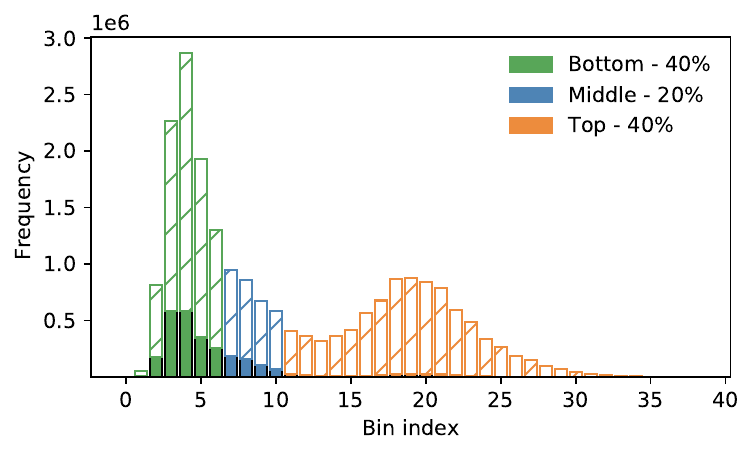}
            \label{subfig:stem_like_distance_distribution}}
	\subfigure[Finish-Specific Embeddings in STEM]{
		\includegraphics[width=0.45\linewidth]{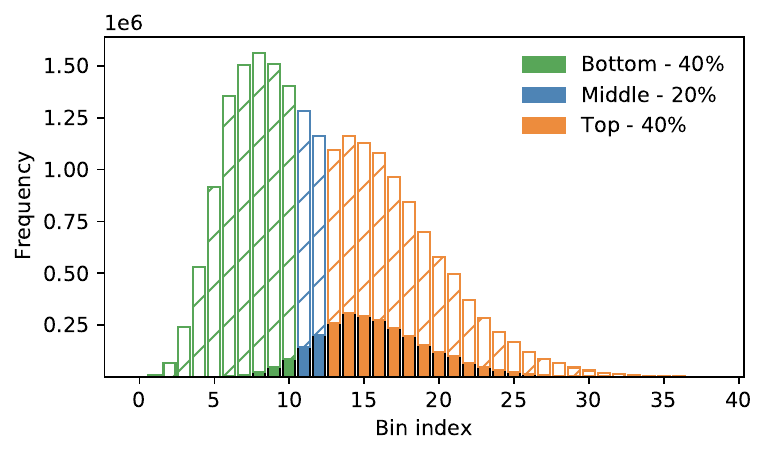}
            \label{subfig:stem_finish_distance_distribution}}
	\subfigure[Shared Embeddings in STEM]{
		\includegraphics[width=0.45\linewidth]{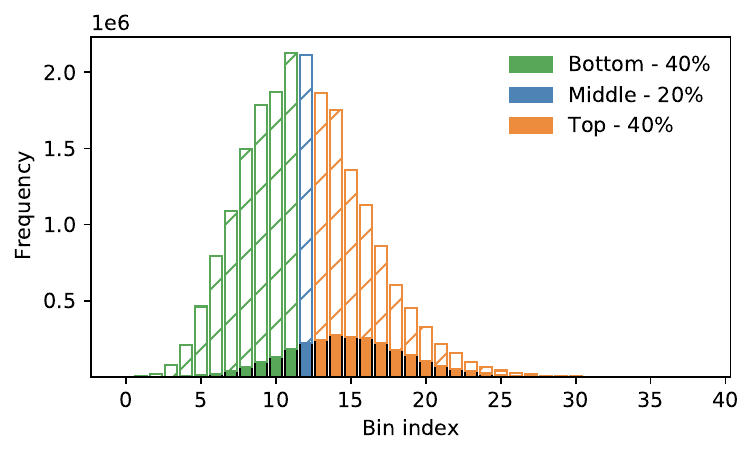}
            \label{subfig:stem_shared_distance_distribution}}
	\caption{Illustration of interest entanglement between tasks in single-embedding based MTL models and disentanglement in STEM. It shows the distance distribution of the contradictory user-item pair set $S$ (with solid color) as well as the whole user-item pair set (with slash lines) regarding the single task Like (a) and Finish  embedding (b), the PLE shared-embedding (c), and the Like (d) and Finish-specific (e) embedding and shared embedding  (f) in STEM.}
	\label{fig: embedding_distance_analysis}
\end{figure}

\begin{figure*}[h]
    \includegraphics[width=0.98\textwidth]{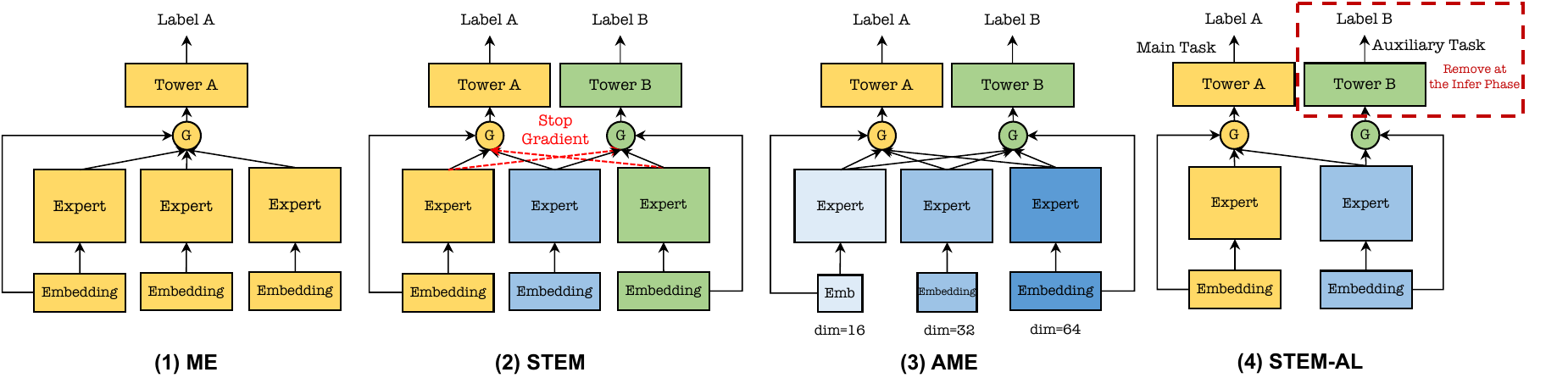}
    \caption{Architecture illustration of various paradigms. 
    Multi-Embedding (ME) is for single-task learning and doesn't disentangle representations. 
    Shared and Task-specific Embedding (STEM) and Asymmetric Multi-Embedding (AME) are both for multi-task learning. STEM disentangles representations via task-specific embeddings, while AME achieves disentanglement through learning multiple embedding tables with different embedding sizes. 
    STEM for Auxiliary Learning (STEM-AL) is for auxiliary learning, which learns task-specific embedding for the main task and a shared embedding updated by multiple tasks.}
    \label{fig:stem_archi}
\end{figure*}

User responses in ads recommender systems are driven by their interests under a specific task or scenario.
Recently, there has been a trend to train multiple tasks or scenarios together to leverage the information from more tasks/scenarios to enhance prediction accuracy.
However, existing work, \textit{e.g.}, MMoE~\cite{MMoE2018} and PLE~\cite{PLE2020}, mainly employ a shared-embedding paradigm~\cite{SharedBottom1997, MMoE2018, MT-FwFM2019, PLE2020, AdaTask2023}, learning one embedding representation for each user and ad.
This leads to a risk of entangling the learned embedding by the possibly contradictory user interests from various tasks or scenarios~\cite{STEM2023}, resulting in \emph{negative transfer}.


Fig.~\ref{fig: embedding_distance_analysis} demonstrates the entanglement of user interest in the public TikTok dataset, which consists of two tasks: Like and Finish.
We select a set of contradictory user-item pairs $S$ whose Euclidean distance is among the bottom-40\% regarding single task Like embedding (Fig.~\ref{subfig:st_like_distance_distribution}) and among the top-40\% regarding single task Finish embedding (Fig.~\ref{subfig:st_finish_distance_distribution}).
We plot the distance distribution of these pairs regarding the shared embedding from PLE in Fig.~\ref{subfig:ple_distance_distribution} and observe that PLE learns large distances for most of them, which is similar to the distribution of single task Finish, while contradictory to that of single task Like.

This section presents two approaches to tackle interest entanglement for multi-task/scenario learning and auxiliary learning.
In the following discussion, we take multi-task learning as an example, and the same principle applies to multi-scenario learning.

\subsection{AME for Multi-Task Learning}

To tackle such an interest entanglement issue, we adopt a Shared and Task-specific EMbedding (STEM) paradigm~\cite{STEM2023}, which \emph{incorporates task-specific embeddings} to learn user's different interest across tasks, along with a shared embedding.
The task-specific embeddings disentangle user and item representation (embeddings) across tasks, making preserving the distinct user interest in different tasks possible, as shown in Fig.~\ref{subfig:stem_like_distance_distribution} and Fig.~\ref{subfig:stem_finish_distance_distribution}.
We then employ a set of experts, each of which is either task-specific or shared across tasks. 
We propose an All Forward Task-specific Backward gating mechanism~\cite{STEM2023} for task-specific towers so that each task tower receives the forward from all experts, while only backward gradients to its corresponding expert and the shared one.

However, there are many tasks in real-world ads recommendation systems.
For example, each conversion type is usually treated as a task~\cite{MT-FwFM2019} when predicting conversions.
There are usually dozens to hundreds of conversion types, making learning an embedding table for each task infeasible.
Therefore, in practice, we group conversion types into groups and treat each group as a task.
On the other hand, we decouple the number of embedding tables from the number of conversion groups, \textit{i.e.}, we learn a fixed number of embedding tables regardless of the number of groups.
We then rely on the gating mechanism to route between embedding tables and conversion groups.
However, these embedding tables may be entangled with each other due to their symmetry.

To this end, we set different embedding sizes for these embedding tables to disentangle them, leading to an Asymmetric Multi-Embedding paradigm, or AME in short, 
as shown in Figure~\ref{fig:stem_archi}.
These embedding tables are disentangled in the sense that small tasks with fewer data need less model capacity and are routed more to the small-size embedding tables via the gating. 
While the other tasks with more data require larger model capacity and are routed to the large-size embedding tables. 
Other disentangled learning techniques can also be used~\cite{ma2019learning, lin2024disentangled}.

\paragraph{Connections to the Multi-Embedding Paradigm}
Multi-Embedding (ME) paradigm is mainly used for single-task learning to tackle the embedding dimensional collapse. 
In contrast, Shared and Task-specific Embedding (STEM) and Asymmetric Multi-Embedding (AME) are mainly used to disentangle user interest representations across various tasks or scenarios.
We try to use AME for single-task learning (\textit{e.g.}, click prediction), but it brings little additional performance gain upon ME.
Similarly, using ME for multi-task learning leads to Multi-Embedding MMoE (ME-MMoE)~\cite{STEM2023}, which has been proven less effective than STEM~\cite{STEM2023} and AME (in our online test) since its embeddings are symmetric and hence may still be entangled.

\paragraph{Deployment Details}
Our conversion prediction model learns more than $100$ conversion types simultaneously.
We group these conversion types into around 30 towers and adopt the asymmetric multi-embedding (AME) paradigm with three embedding tables of embedding sizes $16$, $32$, and $64$, respectively.
Compared to the single embedding baseline PLE with embedding size $K=64$, AME brings $0.32\%$, $0.24\%$, and $0.48\%$ average AUC lifts for three representative scenarios (Moments, Official Accounts, and News), leading to $4.2\%$, $3.9\%$, and $7.1\%$ GMV lift in our online A/B test. 
In particular, the AUC lifts in small tasks such as \textit{Pay} are $0.35\%$, $0.27\%$, and $0.78\%$, respectively, which are much larger than that on other large tasks.
We also train a PLE model with $K=128$ to study the effect of model capacity, but its performance is even worse than the baseline PLE with $K=64$.

\subsection{STEM-AL for Auxiliary Learning}

In industrial recommenders, sometimes we pay more attention to a main task and want to leverage the signals from other tasks to improve the performance of this main task.
For example, the main task in click prediction is to predict the \emph{convertible click}, which leads to the landing page for further conversions.
Besides this valuable feedback, we also collect users' other behaviors regarding the ad: like, favorite, comment, dislike, and dwell time (on video ads).
We want to enhance the performance of the convertible clicks via  Auxiliary Learning upon these additional tasks.

To prevent these auxiliary tasks from entangling a user's interest in the main task, we follow the STEM paradigm~\cite{STEM2023} and adopt a STEM-based Auxiliary Learning architecture, namely, STEM-AL.
In the following discussion, we'll treat Task A as the main task and Task B as the auxiliary one.
As shown in Figure~\ref{fig:stem_archi}, different from STEM and AME, which pay equal attention to all tasks, STEM-AL treats Task A as the primary one and treats Task B as an auxiliary task to improve the performance of A.
In particular, STEM-AL incorporates two embedding tables and two corresponding interaction experts. 
The first embedding table, referred to as the \emph{main embedding table}, is exclusively used by the tower of the primary task A. 
It is only forwarded to and optimized by the primary task, ensuring that the main task's distinctiveness is preserved without interference from other tasks.
The second embedding table, known as the \emph{shared embedding table}, is forwarded to and optimized by the towers of both tasks. 
This shared embedding table allows Task A to benefit from the knowledge and insights from Task B.
The auxiliary tower is removed during inference; only the main task's tower remains active.

\paragraph{Deployment Details}
We deploy STEM-AL to improve the pCTR in one scenario by samples from other domains.
For example, we take the Applet pCTR as the main task and treat Moments pCTR as the auxiliary task.
By using STEM-AL, the CTR of the main task can be improved by $1.16\%$. 
Further, if we use Moments and Channel pCTR as the auxiliary task, the CTR on Applet can be improved by $2.93\%$.

\section{Model Training}
\label{sec:training}

In this section, we describe several training challenges for the model training in ads recommendation and present our solution.
Commonly, the click or conversion prediction task is formalized as a supervised learning problem with the optimization loss $\mathcal{L} = 1/N\sum_{i=1}^{N}\mathrm{BCE}(y_i, f(x_i))$,
here $y_i$ represents labels with $y_i = 1$ denoting a positive label (\textit{e.g.}, click in pCTR) and $y_i = 0$ denoting a negative label (\textit{e.g.}, non-click in pCTR), $x_i$ represents the input, and $\mathrm{BCE}(y_i, f_i) = -y_i\log\sigma(f_i) - (1 - y_i)\log(1 - \sigma(f_i))$ is the Binary Cross Entropy loss.

\subsection{Gradient Vanishing and Ranking Loss}

Recent work~\cite{twitter, jrc} finds out that incorporating an auxiliary ranking loss with BCE loss has shown substantial performance improvement in online advertising. 
However, the efficacy of this combination form is not fully comprehended. 
We examine its efficiency from a new perspective~\cite{UnderstandingRankingLoss2024}: that negative samples suffer from \emph{gradient vanishing} with only BCE loss when the positive feedback is sparse, such as in our pCTR model, where only 0.1\% to 2\% samples are positive (clicks in pCTR). 
Instead, after combining BCE with the ranking loss, we show empirically and theoretically that the gradients become significantly larger~\cite{UnderstandingRankingLoss2024}.
This leads to a lower BCE loss on both the validation set (indicating better classification ability) and the training set (indicating a better optimization procedure).
We kindly advise readers to refer to~\cite{UnderstandingRankingLoss2024} for more details.

\paragraph{Deployment Details} 

The combination of ranking loss with BCE loss is widely deployed in the Moments and Channel pCTR models, with GMV lifts of 0.57\% and 1.08\%, respectively.
It's also deployed in all pLTV models, with an LTV GMV lift of 5.99\%.
Besides, the prediction bias is also reduced, especially on samples with low prediction scores.

\subsection{Repeated Exposure and Weighted Sampling}

Repetitive exposure, that is, displaying the same or similar ads to users within a short period, can enhance user's perception of specific ads but may also risk harming user experience.
To tackle this, we introduced the \emph{Repetitive Exposure Weight} (REW) module to decrease the prediction score of repeated ads for a given user to reduce their exposure.
The core idea is to assign higher weights to the repeated impressions (negative samples).

Specifically, for each repeated impression with negative label, we assign a weight $w_\text{rep} >= 1$ in the original loss: $\mathcal{L} = \dfrac{1}{N}\sum_{i=1}^{N} w_\text{rep}\cdot\mathrm{BCE}(y_i, f(x_i))$.
It considers both the repeated count as well as recency:  $w_\text{rep} = \alpha\cdot w_{\mathrm{count}} + (1 - \alpha) \cdot w_{\mathrm{recency}}$, where $w_{\mathrm{count}}$ is proportional to the total number of exposure (time decayed) of the same or similar ads to this user, and $ w_{\mathrm{recency}}$ considers the time interval between the last repeated impression and the current time.
Please note that these weights would lead to bias in the whole model since they upweight negative samples of repeated exposure. 
We rectify such bias by involving an additional weight $\mathrm{w}_{debias} = (\sum_{i=1}^{N}(1 - y_i)\cdot w_\text{rep}/\sum_{i=1}^{N}(1 - y_i))$ for all positive samples.

\paragraph{Deployment Details} The REW module is widely deployed in Tencent Official Accounts, News, and Video pCTR models, reducing the percentage of repetitive ads exposure by $14.7\%$, $7.8\%$, and $9.7\%$ respectively.

\subsection{Online Learning}
We train our pCTR and pCVR models with online learning, where samples are populated in seconds.
Online learning for pCVR poses a special challenge due to conversion delay feedback.
There is a lot of work~\cite{chapelle2014modeling, yasui2020feedback} to address it. 
Nevertheless, this method will lead to pronounced model bias due to substantial fluctuations of conversion feedback in our system. 
For instance, certain advertisers may report all previous conversions at uncertain times, resulting in an exceptionally high observed CVR at that moment, while reporting no conversions at other times. 
We propose a dynamic online learning method based on the conversion feedback variance in response to these challenges. 
Specifically, a very small variance means that the observed CVR is close to the history CVR, so we can populate the samples as fast as possible. 
Otherwise, when the variance is large, we will set a waiting time to ensure the stability of conversion arrival and reduce the risk of high bias due to arrival fluctuation.

\paragraph{Deployment Details} 
Our approach has been implemented in various scenarios for Tencent Ads, including pCVR models for Tencent Moments, Channel, and Official accounts, with overall GMV lift of 0.3\%, 1.49\%, 1.14\%, and new ad GMV lift of 2.48\%, 0.8\%, 4.34\% respectively, where the new ads refer to ads that have been online within three days. 
Besides, the bias of new ads in all scenarios has decreased from over 10\% to within 1\%.

\subsection{Exploration with Uncertainty Estimates}
So far, our focus is on enhancing the models' ability to accurately predict click or conversion rates, utilizing these scores to rank ads and maximize exploitation while neglecting the exploration. 
However, extensive research has demonstrated the criticality of striking a balance between exploration and exploitation, particularly for cold-start ads. 
Consequently, we propose adopting a Bayesian perspective for CTR modeling, wherein instead of predicting a single point estimate for CTR, we predict a distribution that incorporates uncertainty estimations.

To achieve this, we introduce a Gaussian process (GP) prior distribution to represent the unknown true CTR function. 
We leverage observed data to obtain predictions and uncertainty estimations from the posterior distribution. 
Combining these uncertainty estimates with well-established bandit algorithms, specifically Thompson Sampling (TS), enables us to manage the exploration-exploitation trade-off and enhance long-term utilities effectively.
Formally,
    \begin{equation}
    \begin{aligned}
        \mathrm{pCTR}_{\mathrm{TS}} &= \sigma(\hat{f})\; \mathrm{where}\; \hat{f}\sim\mathcal{N}(\mu(\mathbf{x}^{\star}), \Sigma(\mathbf{x}^{\star}))
    \end{aligned}
    \end{equation}
    where $\mu(\mathbf{x}^{\star})$ and $\Sigma(\mathbf{x}^{\star})$ denotes the mean and variance of the posterior logit value $f(\mathbf{x} ^{\star})$ for test data point $\mathbf{x} ^{\star}$.
   
\paragraph{Deployment Details} The GP-based model is deployed in Tencent Moments pCTR, with GMV lift of $+1.92\%$.

\section{Analysis Tools}
\label{sec:tools}

In this section, we present several off-the-shelf analysis tools on representation learning to analyze the correlation between features, check whether and to what extent embeddings collapse, and examine the entanglement of user interest. 
We release the analysis code in this repository: https://github.com/junwei-pan/RecScope.

\subsection{Feature Correlation}
\label{subsec:measnre_feature_correlation}

We can measure both \emph{ground-truth} and \emph{learned} feature correlation on particular samples or feature combinations via mutual information~\cite{FwFM2018, TIN2023}.
The ground-truth correlation can be calculated by the mutual information between features $X$ and the user's response(label) $Y$ under certain constraints.
In particular, when handling sequential features, we'd like to measure the semantic-temporal correlation between behaviors with specific categories while at specific position $X_{\text{con}_b}$, and the user's response on targets of specific categories $Y_{\text{con}_t}$. 
After defining the constraints on behaviors and the target, \textit{e.g.}, $\text{con}_b$ and $\text{con}_t$, the correlation can be quantified as

\begin{equation}
    \text{Cor} = \text{MI}(X_{\text{con}_b}, Y_{\text{con}_t})
\end{equation}

For example, we can first select a subset of samples with target category $c_t$, and then for behaviors with category $c_i$ while at various positions $p$ (or with various time intervals), we can quantify the correlation as 

\begin{equation}
    \text{Cor}= \text{MI}(X_{C(X)=c_i \land P(X)=p}, Y_{C(Y)=c_t})
\end{equation}

Using this metric, we do observe both strong semantic correlation, \textit{i.e.}, behaviors belonging to the same category as the target exhibit a stronger correlation than those of other categories, and strong temporal correlation, \textit{i.e.}, there is a compelling correlation decrease from the most recent behaviors to the oldest ones.
Please kindly refer to~\cite{TIN2023} for more details.

\subsection{Embedding Dimensional Collapse}
\label{subsec:measure_embedding_collapse}

Dimensional collapse happens when embedding vectors span in a lower-dimensional subspace.
Following the work of ~\cite{UnderstandingDimensionCollapse2021}, we can measure dimensional collapse by conducting a singular value decomposition (SVD) of the embedding matrix of each field.
In particular, given an embedding matrix of field $i$: $\bm{E}_{i} \in \mathbb{R}^{N_{i} \times K}$, after the SVD $\bm{E}_i = U \Sigma V^*, \Sigma = diag(\sigma^k)$, we can get the singular values $\sigma_k$.
Dimensional collapse happens when some singular values are small.
Besides, we can further quantify the dimensional collapse of an embedding matrix by a new metric: Information Abundance (IA)~\cite{ME2023}, which is defined as the sum of all singular values normalized by the maximum singular value:

\begin{definition}[Information Abundance]
    Consider a matrix $\mE\in\mathbb{R}^{D\times K}$ and its singular value decomposition $\mE=\mU\mSigma\mV=\sum\limits_{k=1}^K\sigma_k\vu_k\vv_k^\top$, then the \emph{information abundance} of $\mE$ is defined as

    \[
    \mathrm{IA}(\mE)=\frac{\|\bm{\sigma}\|_1}{\|\bm{\sigma}\|_\infty},
    \]
\end{definition}


\subsection{Interest Entanglement}
\label{subsec:measure_space_conflict}

User responses in ads recommender systems are driven by many factors behind the users’ decision-making processes.
We can measure such factor entanglement by selecting a set of contradictory user-item pairs $S$ whose embedding distances are large in one task but low in another.
We then plot the distance distribution of $S$ based on: a) embeddings from each single task model, b) the embeddings of a shared-embedding multi-task learning model, \textit{e.g.}, PLE, c) the embeddings of each task as well as the shared embedding in STEM.
An example of such analysis is already shown in Fig.~\ref{fig: embedding_distance_analysis}.


\section{Related Work}

\textbf{Feature Encoding.} 
Modeling sequence of user behaviorss have been widely studied ~\cite{DIN2018, DIEN2019, DSIN2019, BST2019, SIM2020, SASRec2018, GRU4Rec2015, BERT4Rec2019}.
Regarding numerical features, existing work can be categorized into two groups: non-discretization~\cite{PNN2016, Youtube2016, DLRM2019} and discretization~\cite{AutoDis2021, N-ary2022}.
Recently, with huge growth in research on LLM, lots of work has been done on how to utilize embeddings learned from these external pre-trained models in recommendation systems~\cite{hou2022towards, hou2023learning, LCRec2023, CTRL2023, yuan2023go, lin2024clickprompt}.

\noindent \textbf{Feature Interactions and Dimensional Collapse.} There are numerous work on the backbone architecture with explicit or implicit feature interaction, from the shallow models FM~\cite{FM2010}, FFM~\cite{FFM2016}, FwFM~\cite{FwFM2018} and FmFM~\cite{FmFM2021}, to deep models such as Wide \& Deep~\cite{WideAndDeep2016}, DeepFM~\cite{DeepFM2017}, xDeepFM~\cite{xDeepFM2018}, AutoInt~\cite{AutoInt2019}, DCN V2~\cite{DCNv22021} and FinalMLP~\cite{FinalMLP2023}. 
Refer to~\cite{DLRSSurvey2019, CTRReview2022} for a comprehensive survey.

The complete collapse has been widely studied in self-supervised learning (SSL)~\cite{chen2020simple, UnderstandingCLfromAlignmentAndUniformity2020}, and  Mixtures-of-Experts (MoE)~\cite{UnderstandingDimensionCollapse2021}.
On the other hand, dimensional collapse has been studied in SSL~\cite{hua2021feature} and contrastive learning~\cite{UnderstandingDimensionCollapse2021}.

\noindent \textbf{Interest Entanglement under MTL and MDL.}
Negative transfer has been a critical challenge in Multi-Task Learning (MTL) and Multi-Domain Learning (MDL).
Shared-embedding paradigm is widely adopted in either MTL~\cite{SharedBottom1997, MMoE2018, MT-FwFM2019, PLE2020, AdaTask2023} and MDL~\cite{Star2021, HiNet2023, PEPNet2023}.
Disentangled Representation Learning (DRL) aims to identify and disentangle the underlying explanatory factors~\cite{RepresentationLearningPerspective2013} in embeddings and has also been widely used in recommendation~\cite{ma2019learning, wang2022disentangled, DisentangledRepresentationLearning2022, lin2024disentangled}. 

\noindent \textbf{Industry Systems.} 
There are already several works on industrial recommender systems
\cite{Youtube2010, AdClickTrench2013, PracticalFacebook2014, crankshaw2017clipper, lu2017practical, DisplayAdvertisingwithRTB2017, Youtube2016, hstu2024}.
We differ from these works in that we pay more attention to the representation, especially from a dimensional collapse and interest entanglement perspective.

\section{Conclusion}
In this paper, we describe an industry ads recommendation system, paying special attention to the representation learning perspective.
We present how to encode features with inherent priors, as well as practices to tackle the dimensional collapse and interest entanglement issue. 
We also showcase several training techniques and analysis tools.
We hope this work can shed light on the future development of this research area.

\clearpage
\begin{acks}

We want to express our sincere gratitude to the following individuals (alphabetical order) for their invaluable contributions: Chen Cai, Chengfei Cai, Genbao Chen, Rong Chen, Xihua Chen, Junwen Cheng, Xi Cheng, Chang Cui, Chao Deng, Guihe Deng, Huiting Deng, Yiming Deng, Zhixiang Feng, Haijie Gu, Weibo Gu, Chaonan Guo, Xian Hu, Ronggeng Huang, Shudong Huang, Renjie Jiang, Tingyu Jiang, Junfeng Kang, Kai Kang, Weijie Kong, Biao Li, Cong Li, Kaixin Li, Lingling Li, Rui Li, Xiaobo Li, Yi Li, Yuxiong Li, Zhaohua Li, Liwei Lin, Wenbo Liu, Yue Liu, Zijun Liu, Hua Lu, Feiheng Luo, Chao Lv, Lei Mu, Zhen Ouyang, Shuai Ren, Xueyu Shi, Xun Song, Jiayu Sun, Yan Tan, Hui Tang, Henghuan Wang, Hongfa Wang, Shaoying Wang, Xiaochen Wang, Yuan Wang, Shifeng Wen, Gengyu Weng, Haiyang Wu, Qinchen Wu, Ruiqian Wu, Zhengtao Wu, Zhiyuan Wu, Datian Xing, Tengfei Xiong, Sheng Xu, Zeen Xu, Yuekui Yang, Zhaohuan Yang, Changan Ye, Chengguo Yin, Qiufang Ying, Jiulong You, Ming Yue, Difei Zeng, Zijian Zeng, Junjie Zhai, Anran Zhang, Haoran Zhang, Jihong Zhang, Kuo Zhang, Linghan Zhang, Ruifeng Zhang, Shangyu Zhang, Tianjin Zhang, Yaqian Zhang, Wenzhe Zhao, Yufei Zheng, Erheng Zhong, Longsha Zhou and Qi Zhou.

\end{acks}



\bibliographystyle{ACM-Reference-Format}
\bibliography{reference}


\end{document}